\documentclass[conference]{IEEEtran}
\usepackage{cite}

\usepackage{amsmath,amssymb}
\usepackage{braket}
\usepackage[hidelinks]{hyperref}
\usepackage{graphicx}
\usepackage{microtype}

\usepackage{placeins}
\usepackage{quantikz}
\usepackage[capitalise]{cleveref}
\crefname{figure}{Fig.}{Figs.}
\Crefname{figure}{Fig.}{Figs.}
\crefname{table}{Table}{Tables}
\Crefname{table}{Table}{Tables}
\crefformat{equation}{#2(#1)#3}
\Crefformat{equation}{#2(#1)#3}
\crefrangeformat{equation}{#3(#1)#4--#5(#2)#6}
\Crefrangeformat{equation}{#3(#1)#4--#5(#2)#6}

\newcommand{\Uprep}{U_{\mathrm{prep}}}
\newcommand{\HM}{H_M}
\newcommand{\Ulabel}{U_{\mathrm{label}}}

\begin{document}
\bstctlcite{IEEEtranBSTCTL:etal}

\title{Distributed Variational Quantum Optimisation by Entanglement-Selective Transport}

\author{\IEEEauthorblockN{Edric Matwiejew, Pascal Elahi, and Ugo Varetto}
\IEEEauthorblockA{Pawsey Supercomputing Research Centre\\
Perth, WA, Australia}
}

\maketitle

\begin{abstract}
Distributed quantum optimisation is challenging because computing the problem cost function across multiple quantum processors requires non-local gates, which can incur overhead in latency and fidelity. Here we introduce \emph{QESTO}, a distributed variational ansatz for graph-based discrete optimisation that requires only persistent pre-shared Bell pairs for remote operations. Using local operations, it encodes local constraint information in the Bell pairs that is leveraged to produce amplitude transfer towards globally valid distributed solution states. QESTO requires one Bell pair per distributed edge of the problem graph and, after initialisation of the Bell states, uses no non-local gates. On two bounded weighted Wang tile-matching problem ensembles, QESTO achieves stronger convergence to low-cost tilings than equivalently partitioned QAOA with no distributed gates at ansatz depths of two or higher, and exceeds the mean performance of monolithic QAOA at the deepest studied depth in both ensembles. These results suggest that persistent entanglement can support useful variational communication while reducing per-layer non-local gate overhead.
\end{abstract}

\section{Introduction}
\label{sec:intro}

The limitation of near-term quantum processing units (QPUs) to modest qubit counts~\cite{preskill2018} motivates the distribution of quantum computation over multiple units. In such programming models, inter-QPU communication is a significant bottleneck~\cite{cacciapuoti2020}. Typical approaches realise non-local gates via Bell-pair-mediated teleportation primitives~\cite{eisert2000,gottesman1999,yimsiriwattana2005}, which consume Bell pairs at a rate proportional to the number of non-local two-qubit operations and require fresh entanglement for every non-local gate~\cite{cacciapuoti2020,caleffi2022,cuomo2020}.

The overhead of non-local gates poses a challenge for the development of distributed quantum variational optimisation algorithms, as in problems of practical interest, the distributed evaluation of quadratic and higher-order terms in the cost function entails a significant non-local computation. Concretely, the ubiquitous Quantum Approximate Optimisation Algorithm~\cite{farhi2014} employs a hybrid quantum-classical loop featuring an ansatz quantum circuit comprising $p$ layers. In each layer, a \emph{phase-separation unitary} encodes solution costs as a superposition of states that are mapped to the problem search space. Next, a \emph{mixing unitary} drives amplitude between states, during which the phase-encoded information produces interference that is manipulated through variational optimisation to amplify the probability of measuring low-cost solutions. Distributing the ansatz naively thus requires a number of non-local gates that is proportional to the number of layers (or \emph{ansatz depth}).

Recent work on distributed quantum optimisation has largely pursued two directions. Decomposition-based DQAOA variants partition the optimisation problem into sub-problems that are executed locally on multiple QPUs and combined classically, including noise-aware approaches tailored to near-term hardware~\cite{kim2024distributedqaoa,chen2024noiseaware}. Gate-based distributed approaches instead split a single monolithic ansatz across multiple QPUs and require non-local gates~\cite{diadamo2021}. At the systems level, compiler frameworks identify and optimise communication patterns in distributed quantum programs~\cite{wu2022autocomm}, while entanglement-efficient packing schemes reduce the Bell-pair cost of implementing non-local gates~\cite{wu2023entanglement}. Circuit-cutting approaches replace non-local operations with local subcircuits and classical reconstruction, but can be poorly matched to the layered structure of variational ans\"atze because repeated cross-partition interactions increase the number of cuts, and hence the reconstruction overhead, with ansatz depth~\cite{tomesh2023}.

Against this backdrop, we introduce QESTO (Quantum Entanglement-Selective Transport Optimisation), a distributed variational ansatz for graph-structured discrete optimisation problems with edge-label compatibility constraints. The ansatz utilises a mechanism in which constraint information is phase-encoded into persistent Bell pairs and leveraged to produce constraint-aware amplitude transfer that operates entirely in the LOSE (Local Operations Shared Entanglement) regime~\cite{schmid2021postquantum}. This reduces the communication overhead associated with non-local gates while providing a scalable mechanism for transmitting constraint information in the distributed quantum dynamics.

The rest of the manuscript is as follows. In \cref{sec:problem}, we introduce the general form of the classical optimisation problem that QESTO addresses, and the bounded weighted Wang tile-matching problem that is the focus of our numerical benchmarks. In \cref{sec:ansatz}, we present the distributed ansatz. \Cref{sec:results} reports numerical results and compares QESTO with existing methods. We provide our conclusions in \cref{sec:conclusion}.

\section{Problem formulation}
\label{sec:problem}

We consider discrete optimisation problems defined on a \emph{problem graph} $G_P = (V_P, E_P)$ with $|V_P|$ vertices and maximum degree $\Delta$, in which each vertex $v \in V_P$ hosts a discrete variable $x_v$ drawn from a finite set $\Omega_v$, and each edge $e = (u,v) \in E_P$ carries a pairwise compatibility constraint between $x_u$ and $x_v$. For each edge $e \in E_P$, let $\ell_e^{(u)} : \Omega_u \to \{0,1\}$ and $\ell_e^{(v)} : \Omega_v \to \{0,1\}$ denote binary labels that the two vertices present across $e$. For a configuration $\mathbf{x} = (x_v)_{v \in V_P}$, the per-edge mismatch cost is
\begin{equation}
  C_e(x_u, x_v)
  = \lambda_e \bigl(1 - \delta_{\ell_e^{(u)}(x_u),\, \ell_e^{(v)}(x_v)}\bigr),
  \label{eq:paircostgen}
\end{equation}
where $\lambda_e > 0$ is the mismatch penalty weight on edge $e$ and the total problem cost function is
\begin{equation}
  C(\mathbf{x})
  = \sum_{v \in V_P} C_v(x_v)
  + \sum_{e=(u,v) \in E_P} C_e(x_u, x_v),
  \label{eq:costgen}
\end{equation}
where $C_v : \Omega_v \to \mathbb{R}$ is the cost of assigning $x_v$ to vertex $v$.

An NP-hard problem that falls within the scope of \cref{eq:costgen} is \emph{weighted Wang tile-matching}. A regular grid of tiles chosen from a tile set $x_v \in \Omega_v$, with position-dependent placement costs and binary edge labels, is represented by a planar problem graph $G_P$ in which each vertex corresponds to a tile assignment~\cite{wang1961,demaine2007,ansotegui2013}. The task is to minimise the total placement and mismatch costs. For the problem instances considered in this work, each vertex shares the same tile set, so $\Omega_v=\Omega$ and each tile assignment is encoded in a register of width $b=\lceil\log_2|\Omega|\rceil$.

\section{Distributed ansatz}
\label{sec:ansatz}

\begin{figure*}[t]
  \centering
    \scalebox{0.85}{
    \begin{quantikz}[row sep={0.72cm,between origins}, column sep=0.28cm,
                     slice style={draw=black!60, dashed},
                     slice label style={anchor=south, rotate=0, inner sep=1pt, font=\footnotesize}]
      \lstick{$L_1\,/\,n_1$}
        & \gate{\Uprep^{(1)}}
        & \qw
        & \qw \slice{$\ket{\psi_0}$}
        & \gate{U_C^{(1)}(\gamma_r^{(1)})}
        & \gate{U_M^{(1)}(\beta_r^{(1)})}
        & \gate[3,style={inner ysep=-2pt}]{\Ulabel^{(1)}}
        & \qw
        & \qw
        & \qw
        & \gate[3,style={inner ysep=-2pt}]{(\Ulabel^{(1)})^\dagger}
        & \qw \slice{$\times p$}
        & \meter{} \\
      \lstick{$c_{e_1,1}$}
        & \gate{H}
        & \ctrl{3}
        & \qw
        & \qw
        & \qw
        &
        & \gate{R_X(t_r)}
        & \gate{R_Z(\eta_r)}
        & \gate{R_X(t_r)}
        &
        & \qw
        & \qw \\
      \lstick{$c_{e_2,1}$}
        & \gate{H}
        & \qw
        & \ctrl{1}
        & \qw
        & \qw
        &
        & \gate{R_X(t_r)}
        & \gate{R_Z(\eta_r)}
        & \gate{R_X(t_r)}
        &
        & \qw
        & \qw \\
      \lstick{$c_{e_2,2}$}
        & \qw
        & \qw
        & \targ{}
        & \qw
        & \qw
        & \gate[3,style={inner ysep=-2pt}]{\Ulabel^{(2)}}
        & \gate{R_X(t_r)}
        & \gate{R_Z(\eta_r)}
        & \gate{R_X(t_r)}
        & \gate[3,style={inner ysep=-2pt}]{(\Ulabel^{(2)})^\dagger}
        & \qw
        & \qw \\
      \lstick{$c_{e_1,2}$}
        & \qw
        & \targ{}
        & \qw
        & \qw
        & \qw
        &
        & \gate{R_X(t_r)}
        & \gate{R_Z(\eta_r)}
        & \gate{R_X(t_r)}
        &
        & \qw
        & \qw \\
      \lstick{$L_2\,/\,n_2$}
        & \gate{\Uprep^{(2)}}
        & \qw
        & \qw
        & \gate{U_C^{(2)}(\gamma_r^{(2)})}
        & \gate{U_M^{(2)}(\beta_r^{(2)})}
        &
        & \qw
        & \qw
        & \qw
        &
        & \qw
        & \meter{}
    \end{quantikz}
    }
  \caption{Representative depth-$p$ QESTO circuit for two QPUs sharing two distributed edges $e_1,e_2\in E_{\mathrm{dist}}$. Each QPU hosts a local problem register $L_i$ and one communication qubit $c_{e_k,i}$ per incident distributed edge, with one Bell pair prepared across each edge. One ansatz layer applies local cost and mixing unitaries, followed by label encoding, transport, and decoding, as in \cref{eq:layer}.}
  \label{fig:qesto_circuit}
\end{figure*}
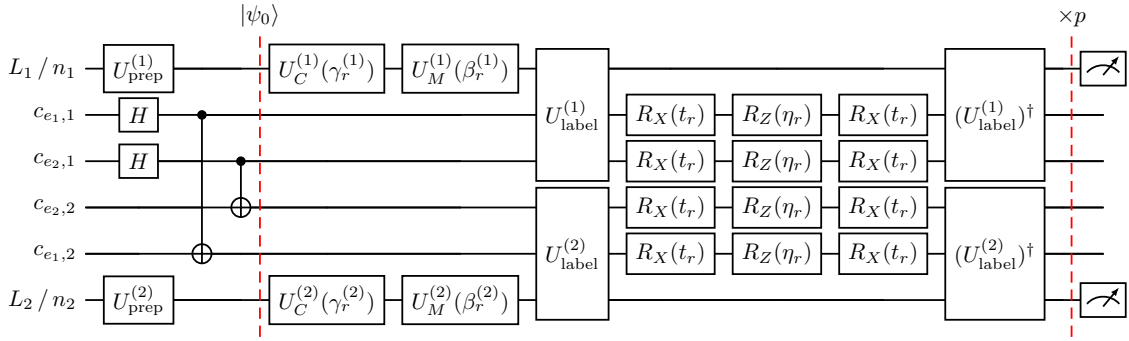

\subsection{Distributed layout and state initialisation}
 
A problem instance is distributed over $m$ QPUs, with each problem-graph vertex assigned to exactly one QPU. We write $V_i$ for the set of problem-graph vertices hosted by QPU~$i$, and $x_i=(x_v)_{v\in V_i}$ for the corresponding local configuration, with valid encoded configuration set $\Omega_i:=\prod_{v\in V_i}\Omega_v$. This partition splits the problem edge set into \emph{local edges} $E_{\mathrm{loc}}$ that are confined to a single QPU and \emph{distributed edges} $E_{\mathrm{dist}}$ that span two QPUs. For each $e\in E_{\mathrm{dist}}$, we denote its QPU endpoints as $\partial e=\{i_e,j_e\}$. If edge $e$ has endpoint $u\in V_i$, we write $\ell_e^{(i)}(x_i):=\ell_e^{(u)}(x_u)$ for the induced QPU-local edge label.

Each QPU~$i$ hosts a \emph{local problem register} $L_i$ with Hilbert space $\mathcal{H}_{L_i} \cong (\mathbb{C}^2)^{\otimes n_i}$, where $n_i = \sum_{v \in V_i}\lceil \log_2 |\Omega_v| \rceil$. The total problem-register width is $n = \sum_{i=1}^{m} n_i$. Each QPU also hosts one \emph{communication qubit} $c_{e,i}$ for every distributed edge $e \in E_{\mathrm{dist}}$ incident on QPU~$i$. The full Hilbert space is
\begin{equation}
  \mathcal{H}
  = \bigotimes_{i=1}^{m}
    \Biggl(\mathcal{H}_{L_i} \otimes \bigotimes_{e \in E_{\mathrm{dist}} :\, i \in \partial e}\!\!\mathbb{C}^2_{c_{e,i}}\Biggr).
  \label{eq:hilbert}
\end{equation}
Each distributed edge is realised by a Bell pair $\ket{\Phi^+} = (\ket{00}+\ket{11})/\sqrt{2}$ shared between $c_{e,i_e}$ and $c_{e,j_e}$, and the joint classical configurations of the problem-graph vertices in $V_i$ are encoded into the computational-basis states of the local problem register $L_i$. 

The initial state of the local problem register is prepared by a state preparation unitary $\Uprep^{(i)}$ as
\begin{equation}
  \Uprep^{(i)} \ket{0^{n_i}}
  = \ket{u_i}
  := \frac{1}{\sqrt{|\Omega_i|}} \sum_{x_i \in \Omega_i} \ket{x_i},
  \label{eq:prep}
\end{equation}
such that the initial joint distributed state of the full system is
\begin{equation}
  \ket{\psi_0}
  = \biggl(\bigotimes_{i=1}^{m} \ket{u_i}_{L_i}\biggr)
    \otimes
    \biggl(\bigotimes_{e \in E_{\mathrm{dist}}} \ket{\Phi^+}_{c_{e,i_e}\, c_{e,j_e}}\biggr).
  \label{eq:initial}
\end{equation}

\subsection{Local phase separation and mixing unitaries}

A \emph{local cost Hamiltonian} collects every term in \cref{eq:costgen} that lives entirely on QPU~$i$,
\begin{equation}
   H_C^{(i)} = \sum_{x_i \in \Omega_i} C^{(i)}(x_i)\, \ket{x_i}\!\bra{x_i}_{L_i},
  \label{eq:cost}
\end{equation}
where
\begin{equation} 
  C^{(i)}(x_i) = \sum_{v \in V_i} C_v(x_v) + \sum_{e=(u,v) \in E_{\mathrm{loc}},\, e \subseteq V_i} C_e(x_u, x_v). 
\end{equation}

Together, the local cost Hamiltonians define the phase-separation unitary,
\begin{equation}
  U_C(\boldsymbol{\gamma}_r)
  = \bigotimes_{i=1}^{m} U_C^{(i)}(\gamma_r^{(i)}),
  \qquad
  U_C^{(i)}(\gamma_r^{(i)})
  = e^{-i\gamma_r^{(i)} H_C^{(i)}},
  \label{eq:phase}
\end{equation}
where we permit one cost-phase angle per QPU $\boldsymbol{\gamma}_r = (\gamma_r^{(1)},\dots,\gamma_r^{(m)}) \in \mathbb{R}^m$.

Together, the local mixing unitaries define
\begin{equation}
  U_M(\boldsymbol{\beta}_r)
  = \prod_{i=1}^{m} U_M^{(i)}(\beta_r^{(i)}),
  \qquad
  U_M^{(i)}(\beta_r^{(i)})
  = e^{-i\beta_r^{(i)} \HM^{(i)}}.
  \label{eq:locmix}
\end{equation}
where $\beta_r^{(i)}$ is a per-QPU mixing parameter and $H_M^{(i)}$ is the local mixing Hamiltonian for QPU $i$. The choice of $H_M^{(i)}$ is flexible and can be tailored to problem-specific requirements~\cite{MarshWang2020,matwiejew2024}.

\subsection{Bell-pair communication layer}

For each distributed edge $e \in E_{\mathrm{dist}}$, each endpoint applies a configuration-controlled $R_Z$ rotation to its communication qubit using a \emph{local edge-labelling unitary}. For endpoint $i\in\partial e$ this is
\begin{equation}
  \Ulabel^{(e,i)}
  = \sum_{x_i \in \Omega_i} \ket{x_i}\!\bra{x_i}_{L_i}
    \otimes R_Z\bigl(\tfrac{\pi}{2} - \pi\ell_e^{(i)}(x_i)\bigr)_{c_{e,i}},
  \label{eq:encode_right}
\end{equation}
We write $\Ulabel^{(i)}$ for the product of these local edge-labelling unitaries over distributed edges incident on QPU~$i$, and $\Ulabel=\prod_{i=1}^{m}\Ulabel^{(i)}$ for their global product. To each communication qubit, we then apply the \emph{transport unitary}
\begin{equation}
  \begin{aligned}
    &U_{\mathrm{trans}}(t_r,\eta_r) \\
    &\quad = \Bigl(\!\prod_{e\in E_{\mathrm{dist}}}\prod_{s\in\partial e} R_X(t_r)_{c_{e,s}}\!\Bigr)
      \Bigl(\!\prod_{e\in E_{\mathrm{dist}}}\prod_{s\in\partial e} R_Z(\eta_r)_{c_{e,s}}\!\Bigr) \\
    &\qquad \times \Bigl(\!\prod_{e\in E_{\mathrm{dist}}}\prod_{s\in\partial e} R_X(t_r)_{c_{e,s}}\!\Bigr),
  \end{aligned}
  \label{eq:transport}
\end{equation}
where $t_r$ and $\eta_r$ are real variational parameters. Finally, the label encoding is uncomputed by $\Ulabel^\dagger$.

\subsection{Ansatz state and objective evaluation}

Altogether, the unitary for one QESTO ansatz layer is
\begin{equation}
  U_{\mathrm{layer}}^{(r)}
  = \Ulabel^\dagger \;
    U_{\mathrm{trans}}(t_r,\eta_r) \;
    \Ulabel \;
    U_M(\boldsymbol{\beta}_r) \;
    U_C(\boldsymbol{\gamma}_r),
  \label{eq:layer}
\end{equation}
and the full depth-$p$ ansatz state is $\ket{\psi_p}= \prod_{r=1}^p U_{\mathrm{layer}}^{(r)} \ket{\psi_0}$.

The variational parameters $\{(\boldsymbol{\gamma}_r,\boldsymbol{\beta}_r,t_r,\eta_r)\}_{r=1}^p$, totalling $(2m+2)p$ scalars, are then tuned to minimise the expected solution cost,
\begin{equation}
  \langle C \rangle_p
  = \sum_{\mathbf{x}} P_p(\mathbf{x})\, C(\mathbf{x}),
  \label{eq:objective}
\end{equation}
where $P_p(\mathbf{x})$ is the marginal probability of measuring distributed configuration $\mathbf{x}$ on the problem registers from $\ket{\psi_p}$ and $C(\mathbf{x})$ is inclusive of cost contributions from all local and distributed edges (exactly as defined in \cref{eq:costgen}). A circuit overview for a 2-QPU QESTO instance with two distributed edges is shown in \cref{fig:qesto_circuit}.

\subsection{Interpretation of the Bell-pair communication layer}

The \emph{Bell-pair communication layer} can be understood geometrically in the Bell basis, with $\ket{\Phi^\pm}=(\ket{00}\pm\ket{11})/\sqrt{2}$ and $\ket{\Psi^\pm}=(\ket{01}\pm\ket{10})/\sqrt{2}$. In this basis, the relevant dynamics occur in two intersecting two-level planes, each of which admits an effective Bloch-sphere description.

The first is the \emph{label-encoding} plane $\mathcal H_Z^{(e)}= \mathrm{span}\{\ket{\Phi^+},\ket{\Phi^-}\}.$ For configuration $(x_{i_e},x_{j_e})$ on the distributed edge $e\in E_{\mathrm{dist}}$, the local edge-labelling unitaries produce the relative phase difference $\varphi_e=\pi\bigl(1-\ell_e^{(i_e)}(x_{i_e})\allowbreak-\ell_e^{(j_e)}(x_{j_e})\bigr)$. Acting on the initial Bell state gives the compatibility-dependent rotation $\ket{\Phi^+}\mapsto \cos(\varphi_e/2)\ket{\Phi^+}\allowbreak-i\sin(\varphi_e/2)\ket{\Phi^-}$.

The second is the \emph{transport-active} plane $\mathcal H_X^{(e)}=\mathrm{span}\{\ket{\Phi^+},\ket{\Psi^+}\},$ in which the generator of the Pauli-X rotation on each communication qubit, $S_X^{(e)} = X_{c_{e,i_e}} + X_{c_{e,j_e}}$, acts as $2X$ on $\mathcal H_X^{(e)}$, leaving the remaining two Bell states invariant.

Together, these mechanisms produce a compatibility-dependent transport operation. For mismatched labels, $\varphi_e\equiv 0 \pmod{2\pi}$, so the label encoding leaves the initial Bell-pair state at $\ket{\Phi^+}$, which intersects with the transport-active plane. By contrast, for matching labels, $\varphi_e\equiv \pi \pmod{2\pi}$, so $\ket{\Phi^+}$ rotates to the \emph{transport-inactive state} $\ket{\Phi^-}$.

Relative to this mechanism, the $\eta$-parameterised Pauli-$Z$ rotation on each communication qubit acts as a tunable rotation that transfers amplitude between $\ket{\Phi^+}$ and the transport-inactive state $\ket{\Phi^-}$. This provides a variational control that can partially undo, enhance, or interpolate the compatibility-dependent transfer induced by the label encoding.

The potential for the overall mechanism to produce constraint-aware amplitude transfer in practice can be illustrated by considering a single distributed edge in the $\ket{\Phi^+}$ state with uniform local costs and identical local mixing unitaries $R_X(\beta_r)$. At depth $p=2$, the parameter choice $(t_1,\eta_1)=(-\pi/2,\pi/2),\allowbreak \beta_1=0,\allowbreak \beta_2=-3\pi/2$ and $(t_2,\eta_2)=(0,0)$ transfers the marginal local-register probability entirely onto the matching-label subspace. For independent single-qubit edge registers with uniform local costs and separable local mixing, the transport unitary factorises over distributed edges, so this compatibility-selective mechanism extends directly to any number of distributed edges. This suggests that the same mechanism can support constraint-aware amplitude transfer when combined with non-trivial local dynamics at low ansatz depth.

\section{Methods}

\label{sec:methods}

We benchmark QESTO against two QAOA references on ensembles of bounded weighted Wang tile-matching instances~\cite {tyburec2023bounded}. The first, which we denote \emph{subgraph-QAOA}, follows the same problem partitioning and problem cost function $C(\mathbf{x})$ as QESTO, with no quantum communication between QPUs. The second is \emph{monolithic QAOA}, which acts on the full $n$-qubit problem register and applies the complete placement-plus-mismatch cost in its phase-separation unitary. We focus on whether each ansatz, under optimised variational parameters, can concentrate probability on low-cost solutions. This isolates the efficacy of the ansatz itself, a necessary condition for developing measurement-efficient parameterisation schemes~\cite{bennett2026benchmarking,montanez2025toward}.

\begin{figure}[t]
  \centering
  \includegraphics[width=0.95\columnwidth]{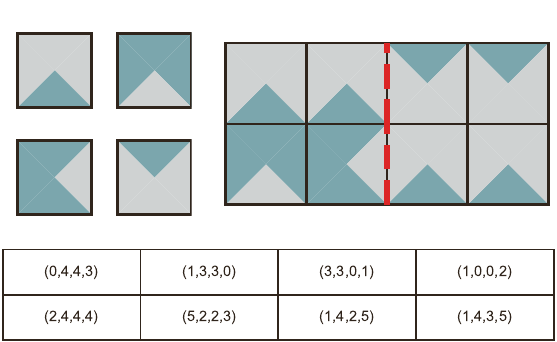}
  \caption{Example bounded weighted Wang tile-matching instance from the $2{\times}4$ ensemble. The top-left panel shows the tile catalogue, the top-right panel shows a globally optimal solution, and the bottom panel gives the position-dependent tile-assignment costs, with each tuple ordered clockwise from the upper-left catalogue tile. Blue and grey wedges encode the binary edge labels. The vertical red dashed line marks the inter-QPU boundary for optimisation by QESTO and subgraph-QAOA.}
  \label{fig:wang_instance}
\end{figure}

The first problem ensemble uses a $2{\times}2$ rectangular problem graph with $|V_P|=4$ vertices, partitioned column-wise over two QPUs, such that there are two parallel distributed edges. The second ensemble uses a $2{\times}4$ rectangular problem graph with $|V_P|=8$ vertices, partitioned column-wise into two $2{\times}2$ QPU-local subproblems. This preserves the number of QPUs and distributed edges but doubles the per-QPU local problem register size. In both ensembles, each vertex is assigned one of four Wang tiles ($|\Omega_v|=4$, two tile-encoding qubits per vertex), with each ensemble comprising $20$ instances with random tile colour sets and per-vertex placement costs. We use a uniform mismatch penalty $\lambda_e=\lambda$ on every edge, with $\lambda$ set to the median of the placement costs for that instance. Every instance is filtered to be a non-trivial distributed problem, meaning that the optimal configuration for each QPU-local subproblem yields a strictly suboptimal global solution. For the benchmark circuits, we follow the mixing unitary design heuristic of quantum walk-based optimisation, choosing $H_M^{(i)}$ to be the adjacency matrix of the Hamming graph $H(|V_i|,|\Omega|)$ over the local tile assignments on $V_i$, in which two configurations are adjacent if they differ in the tile choice at exactly one vertex~\cite{MarshWang2020,matwiejew2024}. To reduce the classical compute footprint associated with circuit simulation, the diagonal local phase-separation and local edge-labelling unitaries were implemented via Walsh--Hadamard decomposition~\cite{welch2014}, achieving ancilla-free implementations at the expense of higher circuit depth. We summarise the benchmark configurations, including the search-space sizes, qubit counts, and parameter counts for each method in \cref{tab:benchmark_sizes}. An example instance from the $2{\times}4$ ensemble is shown in \cref{fig:wang_instance}.

\begin{table}[t]

  \centering

  \caption{Benchmark Configurations}

  \label{tab:benchmark_sizes}

{\scriptsize

  \setlength{\tabcolsep}{2pt}

  \begin{tabular}{ccccc}

    \hline

    Grid Size & Ansatz & Total qubits & Search space & Parameter count\\

    \hline

    $2{\times}2$ & Monolithic QAOA & $8$ & $256$ & $2p$ \\

    $2{\times}2$ & Subgraph-QAOA & $8$ ($4$/QPU) & $256$ & $4p$ \\

    $2{\times}2$ & QESTO & $12$ ($6$/QPU) & $256$ & $6p$ \\

    $2{\times}4$ & Monolithic QAOA & $16$ & $65{,}536$ & $2p$ \\

    $2{\times}4$ & Subgraph-QAOA & $16$ ($8$/QPU) & $65{,}536$ & $4p$ \\

    $2{\times}4$ & QESTO & $20$ ($10$/QPU) & $65{,}536$ & $6p$ \\

    \hline

  \end{tabular}

}%

\end{table}

The $2{\times}2$ and $2{\times}4$ ensembles are evaluated at depths $p \in \{1,\ldots,5\}$ and $p \in \{1,\ldots,8\}$, respectively. At each depth, all three methods perform $8$ independent restarts of the limited-memory Broyden-Fletcher-Goldfarb-Shanno algorithm with bound constraints (L-BFGS-B)~\cite{byrd1995}, each with a maximum of $2000$ iterations. Initial values are drawn uniformly and bounded to the interval of $[-3\pi/4,3\pi/4]$ for the QESTO transport parameter $t_r$, and $[-\pi,\pi]$ otherwise. For QESTO at depth $p=1$, $(t_1,\eta_1)$ is fixed to $(0,0)$ because, without a subsequent local mixing unitary, the transport unitary does not change the measured marginals of the local problem registers. For $p \ge 2$, all methods use prefix warm starts from the best depth-$(p{-}1)$ parameters, sampling only parameters for the new layer. All circuits are evaluated by noiseless statevector simulation via Qiskit~Aer~\cite{qiskit2024} (Python 3.12, Qiskit 2.4, Qiskit~Aer 0.17, SciPy~\cite{2020SciPy-NMeth} 1.17). Each simulation ran on a NVIDIA GraceHopper GH200 superchip on the ``Setonix'' system at the Pawsey Supercomputing Research Centre.

We choose as the primary figure of merit the \emph{normalised optimality gap}
\begin{equation}
  g \;=\; \frac{\langle C \rangle_p - C^{*}}{C_{\mathrm{ind}} - C^{*}},
  \label{eq:normgap}
\end{equation}
where $\langle C \rangle_p$ is the variationally-optimised expected cost \cref{eq:objective}, $C^{*}$ is the global optimum, and $C_{\mathrm{ind}}$ is the cost obtained by solving each QPU-local subproblem independently and combining the resulting assignments. With this normalisation, $g=0$ corresponds to complete convergence to a global optimum and $g=1$ corresponds to matching the independent-optimum baseline. For each combination of method, instance and depth, we report the minimum $\langle C \rangle_p$ out of the eight restarts. We also report the \emph{feasible probability}, defined as the probability of sampling a tiling with no mismatching edges.

To examine how each ansatz redistributes probability mass across the low-cost region of the solution space, we also report the \emph{cost-rank distribution}. For each instance, we order the distinct classical cost values from lowest to highest and compute the mean measurement probability across all assignments that share the same cost rank. We compare these distributions with those obtained from a uniform random sample of the solution space.

\begin{figure}[t]
  \centering
  \includegraphics[width=0.80\columnwidth]{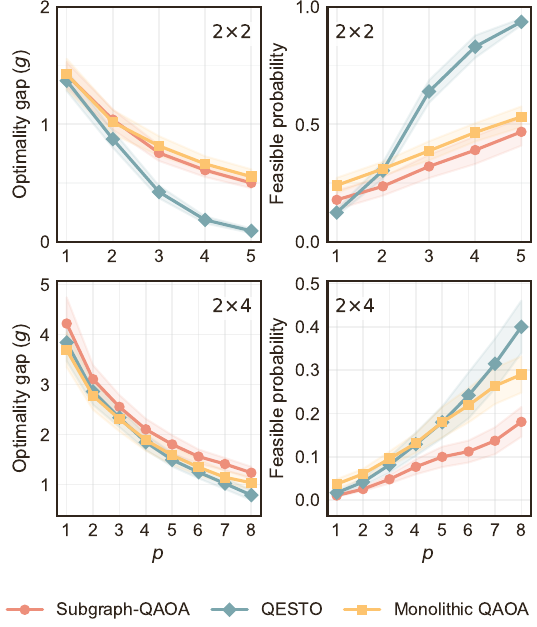}
  \caption{QESTO vs.\ subgraph-QAOA and monolithic QAOA on the $2{\times}2$ (top) and $2{\times}4$ (bottom) layouts. The left panels show the mean normalised optimality gap $g$ (see \cref{eq:normgap}) vs.\ depth $p$, and the right panels show the mean feasible probability vs.\ $p$; shaded bands are $\pm 1$ SEM across instances. All points use $20$ instances.}
  \label{fig:wang_summary}
\end{figure}

\section{Results and discussion}
\label{sec:results}

\subsection{Numerical}
\label{sec:results:numerical}

\Cref{fig:wang_summary} illustrates ansatz performance on the two problem ensembles. On the $2{\times}2$ ensemble, the normalised optimality gaps of the three ans\"atze are close at $p=1$, but from $p=2$ onward separate sharply. QESTO's mean gap drops to $0.87$, $0.43$, $0.19$, and $0.09$ at $p=2,3,4,5$, while subgraph-QAOA reaches only $0.50$ and monolithic QAOA reaches $0.56$ at $p=5$. The right panel shows the same separation in feasible probability. At $p=5$, QESTO's mean feasible probability is $0.936$, while subgraph-QAOA and monolithic QAOA reach mean feasible probabilities of only $0.468$ and $0.532$, respectively, at the same depth.

On the $2{\times}4$ ensemble QESTO's mean normalised gap drops from $3.84$ at $p=1$ to $0.79$ at $p=8$, against $4.22 \to 1.24$ for subgraph-QAOA and $3.69 \to 1.04$ for monolithic QAOA. At $p \geq 2$, QESTO retains a clear advantage over subgraph-QAOA and outperforms monolithic QAOA at $p=8$. A similar trend is present in the mean feasible probability, which at $p=8$ is $0.400$ for QESTO, $0.181$ for subgraph-QAOA, and $0.290$ for monolithic QAOA.

\begin{figure}[t]
  \centering
  \includegraphics[width=0.80\columnwidth]{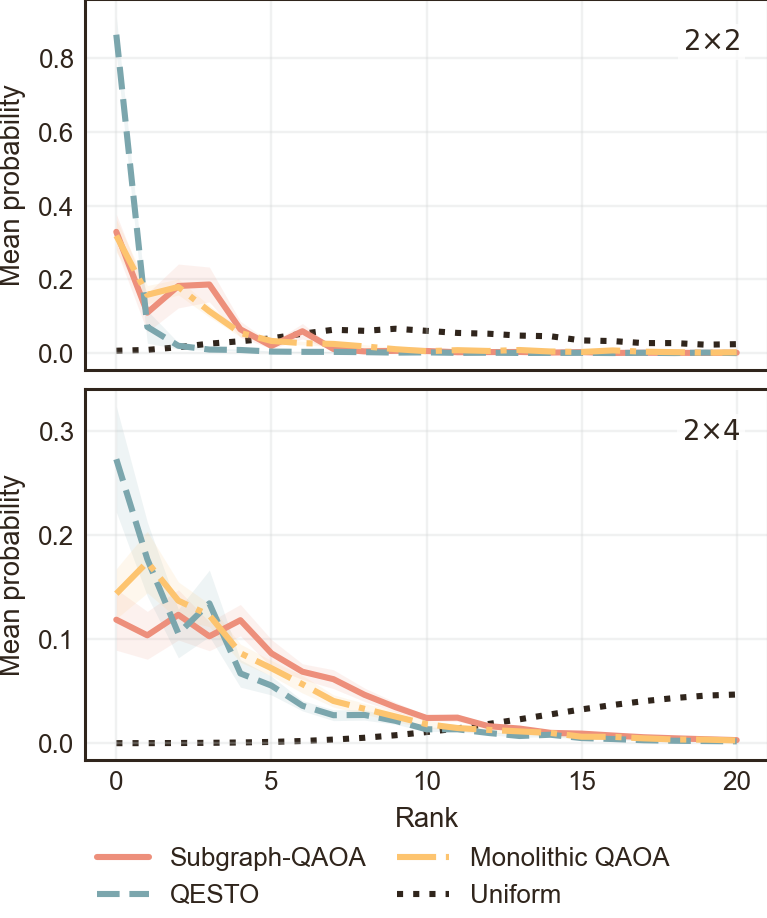}
  \caption{Mean cost-rank distributions for the $2{\times}2$ ensemble at depth $p=5$ (top) and the $2{\times}4$ ensemble at depth $p=8$ (bottom). Rank $0$ denotes the globally optimal cost level, and higher ranks denote successively larger distinct classical cost values. Curves show the mean probability mass across instances; shaded bands are $\pm 1$ SEM. The black dotted line shows the distribution obtained by uniform random sampling of the solution space. All points use $20$ instances.}
  \label{fig:rank_cost_distribution}
\end{figure}

\Cref{fig:rank_cost_distribution} shows how the optimised probability mass is distributed across low-cost ranks. In both ensembles, all three ans\"atze shift substantial probability toward low-rank cost levels relative to uniform sampling, but QESTO produces the strongest concentration near the optimum. On the $2{\times}2$ ensemble at $p=5$, QESTO places a mean probability of $0.863$ on rank $0$, compared with $0.329$ for subgraph-QAOA and $0.318$ for monolithic QAOA. The same trend persists on the harder $2{\times}4$ ensemble at $p=8$, with QESTO placing more probability on rank $0$ and across the lowest few cost ranks than either reference ansatz. This indicates that QESTO's advantage is not confined to improved feasible-solution probability, but extends to more effective redistribution of amplitude toward the low-cost region of the problem search space.

The trends present in \cref{fig:wang_summary} are consistent with the Bell-basis mechanism described in \cref{sec:ansatz}. At $p=1$, the Bell-pair communication layer does not influence the marginal distribution, while from $p\ge 2$ local mixing followed by the transport unitary can redistribute amplitude according to distributed-edge label compatibility. This produces the clearest separation on the $2{\times}2$ ensemble, where each QPU contains only two vertices and matching the two distributed edges is a dominant part of the optimisation landscape. On the $2{\times}4$ ensemble, the number of distributed edges is unchanged, but the per-QPU search space is larger, so while QESTO still improves over subgraph-QAOA, the magnitude of the effect is diluted. Nevertheless, across all studied depths on both ensembles, QESTO matches or exceeds the monolithic QAOA in mean normalised optimality gap. This indicates that QESTO can recover much of the value of the full cross-boundary cost phases without introducing per-layer non-local gates, even at low ansatz depth.

\subsection{Comparison with gate-teleported QAOA}
\label{sec:results:related}

For the Wang tile-matching instances studied here, reversible evaluation of the problem cost function sets the leading local scaling shared by a gate-teleported implementation of the monolithic QAOA and QESTO. With each tile assignment encoded in a $b$-qubit register, a lookup implemented by parallel equality-test blocks over an $|\Omega|$-entry table requires $\mathcal{O}(|\Omega|b)$ two-qubit gates, depth $\mathcal{O}(b)$, and $\mathcal{O}(|\Omega|b)$ work ancillas~\cite{barenco1995}. Since each of the $|V_P|$ tile registers contributes one placement lookup and at most $\Delta$ incident edge-label lookups, reversible cost evaluation over $p$ layers requires $\mathcal{O}(p|V_P|(1+\Delta)|\Omega|\log|\Omega|)$ local two-qubit gates. Computing these mismatch checks in $\mathcal{O}(\Delta)$ rounds of disjoint edges gives depth $\mathcal{O}(p(1+\Delta)\log|\Omega|)$ with $\mathcal{O}(|V_P||\Omega|\log|\Omega|)$ reusable work ancillas.\footnote{The local mixing unitary contributes $\mathcal{O}(p|V_P||\Omega|)$ additional two-qubit gates and at most $\mathcal{O}(|V_P|)$ reusable work ancillas~\cite{matwiejew2024}.} For each distributed-edge endpoint, the QESTO label-encoding and decoding unitaries each add one edge-label lookup and one controlled single-qubit rotation, preserving the same leading local scaling.

Gate-teleported QAOA requires Bell-pair consumption and non-local two-qubit operations, which, in the phase-separation unitary alone, scale as $\mathcal{O}(p|E_{\mathrm{dist}}|)$~\cite{eisert2000}. By contrast, QESTO requires only $|E_{\mathrm{dist}}|$ persistent Bell pairs. While this Bell-pair resource advantage requires the communication qubits to remain coherent throughout the depth-$p$ circuit, current networked demonstrations of remote entanglement-based operations are slower and lower-fidelity than local two-qubit gates~\cite{main2025,saha2025}. As such, maintaining QESTO's persistent Bell-pair coherence may be easier than repeatedly executing non-local gates in a fully distributed ansatz on near-term distributed quantum systems.

\FloatBarrier
\section{Conclusion}
\label{sec:conclusion}

We have introduced QESTO, a distributed variational ansatz for graph-based discrete optimisation problems with constraints. The ansatz uses one pre-shared Bell pair per distributed edge and, after Bell-pair initialisation, requires only local operations. Its key mechanism is a Bell-pair communication layer in which local label encodings imprint distributed constraint information into the Bell basis, and local transport rotations convert this information into compatibility-dependent amplitude redistribution across the partitioned search space.

In numerical simulations on two bounded weighted Wang tile-matching problem ensembles, QESTO achieves stronger convergence to low-cost solutions and a higher probability of measuring solutions without edge mismatches than equivalently partitioned QAOA with no distributed gates at ansatz depths of two and higher, and exceeds the performance of the non-distributed (monolithic) QAOA at the deepest studied depth in both ensembles. This indicates that persistent Bell-pair communication can recover much of the value of a fully distributed approach without introducing per-layer non-local gates, which is likely advantageous on platforms where local gates have higher fidelity and lower execution time than their remote counterparts.

The results obtained focus on the behaviour of the ansatz under ideal conditions with optimised parameters, leaving its study under realistic noise models and measurement overhead for future work. To the latter end, we note that recent work has achieved success in reducing variational optimisation overhead via depth-independent parameterisation schemes and suggest generalising these to distributed ans\"atze as a key area for future research~\cite{bennett2026benchmarking,montanez2025toward}. Other avenues include applying QESTO to larger graphs, additional problem classes, and extending the ansatz to non-binary distributed constraints.

\section*{Acknowledgements}
This work was supported by resources provided by the Pawsey Supercomputing Research Centre (\url{https://doi.org/10.48569/18sb-8s43}), with funding from the Australian Government and the Government of Western Australia. The Pawsey Supercomputing Research Centre's Quantum Supercomputing Innovation Hub and this work were made possible by a grant from the Australian Government through the National Collaborative Research Infrastructure Strategy (NCRIS).

\bibliographystyle{IEEEtran}
\bibliography{references}

\end{document}